\documentclass[aps,prb,twocolumn,superscriptaddress,10pt]{revtex4-1}

\usepackage{graphicx}
\usepackage{amsmath}
\usepackage{amssymb}
\usepackage{amsfonts}
\usepackage{bm}
\usepackage{bbm}
\usepackage[mathcal]{euscript}

\usepackage{color}

\newcommand{\eq}[1]{Eq.~(\ref{#1})}

\begin{document}
\title{Origin of non-linear piezoelectricity in III-V semiconductors:
Internal strain and bond ionicity from hybrid-functional density
functional theory}

\author{Miguel~A. Caro}
\email{mcaroba@gmail.com}
\affiliation{Photonics Theory Group,
Tyndall National Institute, Dyke Parade, Cork, Ireland}
\affiliation{Department of Physics, University College Cork, Cork, Ireland}
\affiliation{Department of Electrical Enegineering and Automation,
Aalto University, Espoo, Finland}
\affiliation{COMP Centre of Excellence in Computational Nanoscience,
Department of Applied Physics, Aalto University, Espoo, Finland}
\author{Stefan Schulz}
\affiliation{Photonics Theory Group,
Tyndall National Institute, Dyke Parade, Cork, Ireland}
\author{Eoin~P. O'Reilly}
\affiliation{Photonics Theory Group,
Tyndall National Institute, Dyke Parade, Cork, Ireland}
\affiliation{Department of Physics, University College Cork, Cork, Ireland}

\begin{abstract}
We derive first- and second-order piezoelectric coefficients for the
zinc-blende III-V semiconductors,
\{Al,Ga,In\}-\{N,P,As,Sb\}. The results are
obtained within the Heyd-Scuseria-Ernzerhof
hybrid-functional approach in the framework of density functional
theory and the Berry-phase theory of electric polarization.
To achieve a meaningful interpretation of the results,
we build an intuitive phenomenological model based
on the description of internal strain and the dynamics of the electronic
charge centers. We discuss
in detail first- and second-order internal strain effects, together with
strain-induced changes in ionicity.
This analysis reveals that the relatively large importance
in the III-Vs of non-linear piezoelectric effects compared to the
linear ones arises because of a delicate balance between the
ionic polarization contribution due to
internal strain relaxation effects, and the contribution due to the
electronic charge redistribution induced by macroscopic and internal strain.
\end{abstract}

\date{\today}

\maketitle

\section{Introduction}

Non-linear piezoelectricity and its importance in III-V semiconductors
and nanostructures has been a topic of intense debate over the last few
years.~\cite{bester_2006,bester_2006_b,migliorato_2006,garg_2009,beya_2011}
The existence of non-negligible second-order piezoelectricity in GaAs and
InAs was first proposed by Bester \textit{et al.}~\cite{bester_2006}
in the context of linear-response density functional theory (DFT).
It was soon revisited semi-empirically
by Migliorato \textit{et al.}~\cite{migliorato_2006}
using Harrison's bond orbital model and a strain-dependent Kleinman
parameter $\zeta$ calculated from DFT.
More recently, following Bester's methodology, Beya-Wakata
\textit{et al.}~\cite{beya_2011}
have extended the calculation of first- and second-order coefficients to
other zinc-blende (ZB) III-V binaries.
Although failing for lower band gap materials due to the intrinsic
limitations of
the local density approximation (LDA),~\cite{perdew_1981}
the approach by Bester \textit{et al}.~\cite{bester_2006} provides, for the
higher energy gap III-V materials,
\textit{ab initio} parameters without need to fit to experiment.
Given that measuring piezoelectric (PZ) constants accurately is
complicated, in particular when possibly large
second-order effects are superimposed to the linear ones, the importance
of theoretical calculations cannot be overstated. It was indeed found by
Beya-Wakata \textit{et al.},~\cite{beya_2011} and is confirmed in the
present work with the use of an improved energy functional, that
second-order effects are large for the family of ZB III-V materials,
in some cases dominating over first-order piezoelectricity even for small
strains (e.g. AlP, AlAs and InP).
However, it remains unclear and hidden beneath the numbers what is
the physical origin of this large second-order
PZ effect, and what is the reason for the
apparent lack of trends in the evolution of the coefficients for
the different compounds.~\cite{beya_2011}
We try to address this question here by looking separately at the different
factors which contribute to the PZ coefficients, investigating
separately the linear and nonlinear contributions to the Kleinman
parameter and to the material electronic response, as well as noting
the influence of volume effects on the PZ coefficient values.

The form of the second-order PZ polarization tensor for the
ZB lattice, along with all the remaining PZ crystal
classes, has been established by Grimmer based on
symmetry arguments.~\cite{grimmer_2007} The full expression for the
PZ vector of a ZB lattice, including first-
and second-order effects, reads:~\cite{beya_2011}
\begin{align}
\textbf{P} \equiv
&
e_{14}
\left(
\begin{array}{c}
\epsilon_4 \\
\epsilon_5 \\
\epsilon_6 \\
\end{array}
\right)
+ B_{114}
\left(
\begin{array}{c}
\epsilon_1 \epsilon_4 \\
\epsilon_2 \epsilon_5 \\
\epsilon_3 \epsilon_6 \\
\end{array}
\right)
\nonumber \\
&
+ B_{124}
\left(
\begin{array}{c}
(\epsilon_2 + \epsilon_3 ) \epsilon_4 \\
(\epsilon_1 + \epsilon_3 ) \epsilon_5 \\
(\epsilon_1 + \epsilon_2 ) \epsilon_6 \\
\end{array}
\right)
+ B_{156}
\left(
\begin{array}{c}
\epsilon_5 \epsilon_6 \\
\epsilon_4 \epsilon_6 \\
\epsilon_4 \epsilon_5 \\
\end{array}
\right),
\label{01}
\end{align}
where $e_{14}$ is the linear PZ coefficient, while $B_{114}$,
$B_{124}$ and $B_{156}$ are the second-order PZ coefficients.
The strain components are denoted by $\epsilon_j$. We follow standard
Voigt notation for all the symbols.

Previous analysis by Beya-Wakata \textit{et al.},~\cite{beya_2011}
found that $B_{114}$ and
$B_{124}$ show no obvious trend with compound. We note that the
two terms in the first ($x$-) component of the polarization vector [\eq{01}]
involving these two second-order PZ coefficients are proportional
to $\epsilon_1$ and  $\epsilon_2 + \epsilon_3$, respectively.
However, it is often more useful and intuitive to describe the response
of a material in terms of the hydrostatic strain
$\text{Tr}(\epsilon) = \epsilon_1 + \epsilon_2 + \epsilon_3$ and the
biaxial strain $\epsilon_{\text{b},i} = (3 \epsilon_i -
\text{Tr}(\epsilon))/2$. Equation~(\ref{01}) can then be rewritten
as~\cite{schulz_2011}
\begin{align}
\textbf{P} \equiv
&
e_{14}
\left(
\begin{array}{c}
\epsilon_4 \\
\epsilon_5 \\
\epsilon_6 \\
\end{array}
\right)
+ A_1 \text{Tr}(\epsilon)
\left(
\begin{array}{c}
\epsilon_4 \\
\epsilon_5 \\
\epsilon_6 \\
\end{array}
\right)
\nonumber \\
&
+ A_2
\left(
\begin{array}{c}
\epsilon_{\text{b},1} \epsilon_4 \\
\epsilon_{\text{b},2} \epsilon_5 \\
\epsilon_{\text{b},3} \epsilon_6 \\
\end{array}
\right)
+ B_{156}
\left(
\begin{array}{c}
\epsilon_5 \epsilon_6 \\
\epsilon_4 \epsilon_6 \\
\epsilon_4 \epsilon_5 \\
\end{array}
\right).
\label{24}
\end{align}
with $A_1$ and $A_2$ related to the more usual coefficients as follows:
\begin{align}
A_1 = \frac{B_{114} + 2 B_{124}}{3},
\qquad
A_2 = \frac{2 (B_{114} - B_{124})}{3}.
\end{align}
While trends in $B_{114}$ and $B_{124}$ alone are difficult to 
extract,~\cite{beya_2011}
we find that $A_1$ and $A_2$ tend to decrease and to increase, respectively,
with increasing unit cell volume. These opposing trends then account
for the difficulty to identify any clear trends in $B_{114}$ and $B_{124}$.
Overall, we show that, once the non-linear dependence
on strain of the bond polarity, in terms of charge redistribution
and internal strain effects,
is taken into account, a complete explanation
of the PZ effect, including its non linearity,
can be provided from first principles alone. We present a consistent picture
of ionicity in ZB materials,
drawn from the well-established relation between the dynamics
of Wannier function centers and electric polarization.~\cite{vanderbilt_1993}
Although data on PZ coefficients is somewhat lacking for
ZB semiconductors, more reliable measurements are available for the
urtzite III-Ns. A very recent study of elastic and PZ
properties of WZ GaN by Witczak \textit{et al}.~\cite{witczak_2015}
serves as validation of the methodology employed here, which we already
followed in Refs.~\onlinecite{caro_2012c} and \onlinecite{caro_2013b},
respectively.

Recent work by Tse \textit{et al}.~\cite{tse_2013} relying on the Harrison
model computed the effective PZ coefficient of GaAs and InAs using a
third-order expansion in strain, which corresponds for a fourth-order
expansion for the polarization. However, the authors neglected the
quadratic contributions arising from shear strain (related to the
$B_{156}$ coefficient) which, as we will show,
are of the same order of magnitude as the other second-order contributions.

The paper is organized as follows. We introduce our computational method
in Sec.~\ref{20}. The main results are then presented in Sec.~\ref{21}.
This is followed by an analysis of the results in Sec.~\ref{22}, where we
provide some general considerations in Sec.~\ref{15}, before
discussing the first-order and second-order coefficients
in Secs.~\ref{16} and \ref{17}, respectively. Finally, we summarize our
conclusions in Sec.~\ref{23}. Some additional details of the calculation
method used are presented in the Appendix.

\section{Computational method}\label{20}

Equations~(\ref{01}) and (\ref{24}) portray all the symmetries applicable to
piezoelectricity which are characteristic of the ZB
lattice,~\cite{nye_1985,grimmer_2007} and therefore form the basis of the
target quantities being calculated here.
The DFT calculations for geometry optimization and self-consistent
wave functions
were carried out using the Heyd-Scuseria-Ernzerhof (HSE)
hybrid functional approach~\cite{heyd_2003,heyd_2004}
within the projector augmented-wave (PAW) method~\cite{bloechl_1994,kresse_1999}
as implemented in the \textsc{vasp} package.~\cite{kresse_1996}
The screening parameter $\mu$ was fixed to 0.2 
in our calculations and the mixing parameter
$\alpha$ to 0.25 (these settings correspond to \textsc{vasp}'s
HSE06 version of the HSE functional). The plane-wave cutoff energy was
set to 600~eV. The semicore $d$ states of Ga and In were treated as
valence states. A $\Gamma$-centered configuration
was used for the $k$-point mesh, with a variable number of divisions.
The convergence of the
PZ coefficients was found to be slow with respect to the number of
\textbf{k} points, as described further in the Appendix.
Table~\ref{04} shows the calculated energy gaps and lattice parameters
for all the compounds considered.
The calculations of electric polarization
were performed employing the Berry-phase technique within the context of
the modern theory of polarization,~\cite{king-smith_1993,vanderbilt_1993,
resta_1994,resta_2007} as implemented by Martijn Marsman in
\textsc{vasp}. We have successfully used a similar approach previously
to study internal strain and electric polarization of wurtzite group-III
nitrides.~\cite{caro_2012c,caro_2013b}

\begin{table*}[t]
\caption{Direct band gap at $\Gamma$, indirect gap (when this is smaller
than the direct one) and lattice parameter, for all the binary
ZB III-V semiconductors, calculated from the HSE approach as explained
throughout the text.
A $\Gamma$-centered $10 \times 10 \times 10$ $k$-point sampling is used for
all of these calculations.}
\begin{ruledtabular}
\label{04}
\begin{tabular}{| r | c c c | r | c c c | r | c c c |}
\hline
 & \multicolumn{3}{c |}{\parbox{4.5cm}{Direct gap at $\Gamma$ (eV)}} & \parbox{1.15cm}{\hspace{1cm}}
 & \multicolumn{3}{c |}{\parbox{4.5cm}{Indirect gap (eV)}} & \parbox{1.15cm}{\hspace{1cm}}
 & \multicolumn{3}{c |}{\parbox{4.5cm}{Lattice parameter (\AA)}}
\\
 & \parbox{1.5cm}{Al} & \parbox{1.5cm}{Ga} & \parbox{1.5cm}{In} &
 & \parbox{1.5cm}{Al} & \parbox{1.5cm}{Ga} & \parbox{1.5cm}{In} &
 & \parbox{1.5cm}{Al} & \parbox{1.5cm}{Ga} & \parbox{1.5cm}{In}
\\ \hline
N  & 5.61 & 3.10 & 0.56 & N  & 4.58 &  n/a & n/a & N  & 4.365 & 4.493 & 4.991 \\
P  & 4.21 & 2.81 & 1.43 & P  & 2.31 & 2.33 & n/a & P  & 5.471 & 5.460 & 5.904 \\
As & 2.86 & 1.33 & 0.41 & As & 2.10 &  n/a & n/a & As & 5.687 & 5.686 & 6.116 \\
Sb & 2.23 & 0.75 & 0.31 & Sb & 1.76 &  n/a & n/a & Sb & 6.188 & 6.152 & 6.563
\\ \hline
\end{tabular}
\end{ruledtabular}
\end{table*}

In order to extract the PZ coefficients from the polarization results,
we have fit to second-order polynomials. We have performed calculations
of $P_x$ for 5 different strain branches $\boldsymbol{\epsilon}^{(i)}$,
which read:
\begin{align}
& \boldsymbol{\epsilon}^{(1)} \equiv \left( 0, 0, 0, \beta, \beta, \beta \right)
\quad \rightarrow \quad e_{14} \, \& \, B_{156},
\nonumber \\
& \boldsymbol{\epsilon}^{(2)} \equiv \left( \alpha, 0, 0, \beta, 0, 0 \right)
\quad \rightarrow \quad e_{14} \, \& \, B_{114},
\nonumber \\
& \boldsymbol{\epsilon}^{(3)} \equiv \left( 0, \alpha, 0, \beta, 0, 0 \right)
\quad \rightarrow \quad e_{14} \, \& \, B_{124},
\nonumber \\
& \boldsymbol{\epsilon}^{(4)} \equiv \left( 0, \alpha, \alpha, \beta, 0, 0 \right)
\quad \rightarrow \quad e_{14} \, \& \, B_{124},
\nonumber \\
& \boldsymbol{\epsilon}^{(5)} \equiv \left( \alpha, \alpha, \alpha, \beta, \beta, \beta \right)
\quad \rightarrow \quad e_{14}, \, B_{156} \, \& \, \frac{B_{114}+2 B_{124}}{3}.
\label{19}
\end{align}
Here the relation between
unstrained and strained structures is given by the strain transformation
matrix, which for an arbitrary lattice vector at equilibrium
$\textbf{R}_\alpha \equiv ( R_{\alpha,1}, R_{\alpha,2}, R_{\alpha,3} )$
is given by
\begin{align}
\textbf{R}_\alpha' =
\left(
\begin{array}{c c c}
1 + \epsilon_1 & \frac{\epsilon_6}{2} & \frac{\epsilon_5}{2} \\
\frac{\epsilon_6}{2} & 1 + \epsilon_2 & \frac{\epsilon_4}{2} \\
\frac{\epsilon_5}{2} & \frac{\epsilon_4}{2} & 1 + \epsilon_3 \\ 
\end{array}
\right)
\left(
\begin{array}{c}
R_{\alpha,1} \\
R_{\alpha,2} \\
R_{\alpha,3} \\
\end{array}
\right).
\end{align}
We increment the $\alpha$ and $\beta$ parameters in \eq{19}
so that the Voigt components
of the strain tensor follow the relations above and allow to calculate
the corresponding PZ coefficients.
For branch $\boldsymbol{\epsilon}^{(1)}$ 9 data points per compound
and per $k$ mesh were calculated, up to $|\beta| = 0.04$.
For the other branches
(which involve two-dimensional fitting) 25 data points were used,
with $|\alpha|$ and $|\beta|$ up to 0.02 and 0.04, respectively. We
have verified that these strain ranges allow a good quality fitting for both
linear and quadratic terms, as shown in the Appendix. Obviously
this leads to plenty of redundant information which is however useful
in order to check consistency between different calculations. These
consistency checks are important given the convergence issues with respect
to the number of \textbf{k} points highlighted in the Appendix.
They also allow to monitor the effect of symmetry reduction which we have
shown to significantly affect the results of calculated elastic and structural
properties of semiconductors.~\cite{caro_2013}

\begin{table*}[p]
\caption{All the different first- and second-order
PZ coefficients of the ZB III-Vs, split into different contributions.
See text for details.}
\begin{ruledtabular}
\label{03}
\begin{tabular}{| r | c c c | r | c c c | r | c c c |}
%
%
\multicolumn{12}{c}{$e_{14}$ (C/m$^2$)}
\\ \hline
 & \multicolumn{3}{c |}{\parbox{4.5cm}{Ionic part: $e_{14}^\text{ion}$ (C/m$^2$)}} & \parbox{1.15cm}{\hspace{1cm}}
 & \multicolumn{3}{c |}{\parbox{4.5cm}{Electronic part: $e_{14}^\text{ele}$ (C/m$^2$)}} & \parbox{1.15cm}{\hspace{1cm}}
 & \multicolumn{3}{c |}{\parbox{4.5cm}{\textbf{Total}: $e_{14}$ (C/m$^2$)}}
\\
 & \parbox{1.5cm}{Al} & \parbox{1.5cm}{Ga} & \parbox{1.5cm}{In} &
 & \parbox{1.5cm}{Al} & \parbox{1.5cm}{Ga} & \parbox{1.5cm}{In} &
 & \parbox{1.5cm}{Al} & \parbox{1.5cm}{Ga} & \parbox{1.5cm}{In}
\\ \hline
N  & $-$2.266 & $-$2.256 & $-$2.409 & N  & 2.814 & 2.622 & 3.002 & N  &  0.548 &  0.366 &  0.593 \\
P  & $-$1.553 & $-$1.437 & $-$1.503 & P  & 1.568 & 1.316 & 1.518 & P  &  0.014 & $-$0.121 &  0.016 \\
As & $-$1.427 & $-$1.313 & $-$1.371 & As & 1.372 & 1.108 & 1.259 & As & $-$0.055 & $-$0.205 & $-$0.111 \\
Sb & $-$1.236 & $-$1.169 & $-$1.183 & Sb & 1.141 & 0.953 & 1.021 & Sb & $-$0.094 & $-$0.216 & $-$0.161
\\ \hline
%
%
\multicolumn{12}{c}{\shortstack{\vspace{0.25cm} \\ $B_{114}$ (C/m$^2$)}}
\\ \hline
 & \multicolumn{3}{c |}{\parbox{4.5cm}{Ionic part: $B_{114}^\text{ion}$ (C/m$^2$)}} & \parbox{1.15cm}{\hspace{1cm}}
 & \multicolumn{3}{c |}{\parbox{4.5cm}{Electronic part: $B_{114}^\text{ele}$ (C/m$^2$)}} & \parbox{1.15cm}{\hspace{1cm}}
 & \multicolumn{3}{c |}{\parbox{4.5cm}{\textbf{Total}: $B_{114}$ (C/m$^2$)}}
\\
 & \parbox{1.5cm}{Al} & \parbox{1.5cm}{Ga} & \parbox{1.5cm}{In} &
 & \parbox{1.5cm}{Al} & \parbox{1.5cm}{Ga} & \parbox{1.5cm}{In} &
 & \parbox{1.5cm}{Al} & \parbox{1.5cm}{Ga} & \parbox{1.5cm}{In}
\\ \hline
N  & 13.08 & 11.89 & 11.40 & N  & $-$19.89 & $-$17.27 & $-$17.37 & N  & $-$6.81 & $-$5.38 & $-$5.96 \\
P  &  7.06 &  6.36 &  6.05 & P  &  $-$9.08 &  $-$7.59 &  $-$7.59 & P  & $-$2.02 & $-$1.23 & $-$1.54 \\
As &  6.38 &  5.84 &  5.43 & As &  $-$7.99 &  $-$6.30 &  $-$6.60 & As & $-$1.61 & $-$0.99 & $-$1.17 \\
Sb &  5.25 &  4.88 &  4.37 & Sb &  $-$6.01 &  $-$5.18 &  $-$5.00 & Sb & $-$0.76 & $-$0.31 & $-$0.62
\\ \hline
%
%
\multicolumn{12}{c}{\shortstack{\vspace{0.25cm} \\ $B_{124}$ (C/m$^2$)}}
\\ \hline
 & \multicolumn{3}{c |}{\parbox{4.5cm}{Ionic part: $B_{124}^\text{ion}$ (C/m$^2$)}} & \parbox{1.15cm}{\hspace{1cm}}
 & \multicolumn{3}{c |}{\parbox{4.5cm}{Electronic part: $B_{124}^\text{ele}$ (C/m$^2$)}} & \parbox{1.15cm}{\hspace{1cm}}
 & \multicolumn{3}{c |}{\parbox{4.5cm}{\textbf{Total}: $B_{124}$ (C/m$^2$)}}
\\
 & \parbox{1.5cm}{Al} & \parbox{1.5cm}{Ga} & \parbox{1.5cm}{In} &
 & \parbox{1.5cm}{Al} & \parbox{1.5cm}{Ga} & \parbox{1.5cm}{In} &
 & \parbox{1.5cm}{Al} & \parbox{1.5cm}{Ga} & \parbox{1.5cm}{In}
\\ \hline
N  & 9.78 & 11.97 & 9.74 & N  & $-$14.83 & $-$18.70 & $-$16.06 & N  & $-$5.04 & $-$6.73 & $-$6.32 \\
P  & 5.98 &  6.29 & 5.99 & P  &  $-$8.75 &  $-$9.56 &  $-$9.61 & P  & $-$2.76 & $-$3.27 & $-$3.62 \\
As & 5.41 &  5.51 & 5.30 & As &  $-$8.00 &  $-$8.72 &  $-$9.62 & As & $-$2.59 & $-$3.21 & $-$4.31 \\
Sb & 4.38 &  4.29 & 4.31 & Sb &  $-$6.37 &  $-$7.06 &  $-$8.36 & Sb & $-$1.99 & $-$2.77 & $-$4.04
\\ \hline
%
%
\multicolumn{12}{c}{\shortstack{\vspace{0.25cm} \\ $B_{156}$ (C/m$^2$)}}
\\ \hline
 & \multicolumn{3}{c |}{\parbox{4.5cm}{Ionic part: $B_{156}^\text{ion}$ (C/m$^2$)}} & \parbox{1.15cm}{\hspace{1cm}}
 & \multicolumn{3}{c |}{\parbox{4.5cm}{Electronic part: $B_{156}^\text{ele}$ (C/m$^2$)}} & \parbox{1.15cm}{\hspace{1cm}}
 & \multicolumn{3}{c |}{\parbox{4.5cm}{\textbf{Total}: $B_{156}$ (C/m$^2$)}}
\\
 & \parbox{1.5cm}{Al} & \parbox{1.5cm}{Ga} & \parbox{1.5cm}{In} &
 & \parbox{1.5cm}{Al} & \parbox{1.5cm}{Ga} & \parbox{1.5cm}{In} &
 & \parbox{1.5cm}{Al} & \parbox{1.5cm}{Ga} & \parbox{1.5cm}{In}
\\ \hline
N  & 6.32 & 5.35 & 3.56 & N  & $-$10.46 & $-$8.52 & $-$5.57 & N  & $-$4.15 & $-$3.18 & $-$2.00 \\
P  & 2.98 & 3.13 & 2.39 & P  &  $-$4.42 & $-$4.51 & $-$3.41 & P  & $-$1.43 & $-$1.38 & $-$1.02 \\
As & 2.82 & 3.02 & 2.27 & As &  $-$4.14 & $-$4.30 & $-$2.74 & As & $-$1.32 & $-$1.28 & $-$0.46 \\
Sb & 2.15 & 2.41 & 2.02 & Sb &  $-$2.97 & $-$3.11 & $-$2.19 & Sb & $-$0.82 & $-$0.70 & $-$0.16
\\ \hline
%
%
\multicolumn{12}{c}{\shortstack{\vspace{0.25cm} \\ $A_{1}$ (C/m$^2$)}}
\\ \hline
 & \multicolumn{3}{c |}{\parbox{4.5cm}{Ionic part: $A_{1}^\text{ion}$ (C/m$^2$)}} & \parbox{1.15cm}{\hspace{1cm}}
 & \multicolumn{3}{c |}{\parbox{4.5cm}{Electronic part: $A_{1}^\text{ele}$ (C/m$^2$)}} & \parbox{1.15cm}{\hspace{1cm}}
 & \multicolumn{3}{c |}{\parbox{4.5cm}{\textbf{Total}: $A_{1}$ (C/m$^2$)}}
\\
 & \parbox{1.5cm}{Al} & \parbox{1.5cm}{Ga} & \parbox{1.5cm}{In} &
 & \parbox{1.5cm}{Al} & \parbox{1.5cm}{Ga} & \parbox{1.5cm}{In} &
 & \parbox{1.5cm}{Al} & \parbox{1.5cm}{Ga} & \parbox{1.5cm}{In}
\\ \hline
N  & 10.88 & 11.95 & 10.30 & N  & $-$16.52 & $-$18.22 & $-$16.50 & N  & $-$5.63 & $-$6.28 & $-$6.20 \\
P  &  6.34 &  6.31 &  6.01 & P  &  $-$8.86 &  $-$8.90 &  $-$8.94 & P  & $-$2.52 & $-$2.59 & $-$2.93 \\
As &  5.73 &  5.62 &  5.34 & As &  $-$7.99 &  $-$7.91 &  $-$8.62 & As & $-$2.26 & $-$2.47 & $-$3.27 \\
Sb &  4.67 &  4.48 &  4.33 & Sb &  $-$6.25 &  $-$6.44 &  $-$7.24 & Sb & $-$1.58 & $-$1.95 & $-$2.90
\\ \hline
%
%
\multicolumn{12}{c}{\shortstack{\vspace{0.25cm} \\ $A_{2}$ (C/m$^2$)}}
\\ \hline
 & \multicolumn{3}{c |}{\parbox{4.5cm}{Ionic part: $A_{2}^\text{ion}$ (C/m$^2$)}} & \parbox{1.15cm}{\hspace{1cm}}
 & \multicolumn{3}{c |}{\parbox{4.5cm}{Electronic part: $A_{2}^\text{ele}$ (C/m$^2$)}} & \parbox{1.15cm}{\hspace{1cm}}
 & \multicolumn{3}{c |}{\parbox{4.5cm}{\textbf{Total}: $A_{2}$ (C/m$^2$)}}
\\
 & \parbox{1.5cm}{Al} & \parbox{1.5cm}{Ga} & \parbox{1.5cm}{In} &
 & \parbox{1.5cm}{Al} & \parbox{1.5cm}{Ga} & \parbox{1.5cm}{In} &
 & \parbox{1.5cm}{Al} & \parbox{1.5cm}{Ga} & \parbox{1.5cm}{In}
\\ \hline
N  & 2.20 & $-$0.05 & 1.10 & N  & $-$3.38 & 0.96 & $-$0.87 & N  & $-$1.18 & 0.90 & 0.24 \\
P  & 0.72 &  0.05 & 0.04 & P  & $-$0.22 & 1.32 &  1.34 & P  &  0.49 & 1.36 & 1.38 \\
As & 0.64 &  0.22 & 0.09 & As &  0.01 & 1.61 &  2.02 & As &  0.65 & 1.48 & 2.09 \\
Sb & 0.58 &  0.40 & 0.04 & Sb &  0.24 & 1.25 &  2.24 & Sb &  0.82 & 1.64 & 2.27
\\ \hline
\end{tabular}
\end{ruledtabular}
\end{table*}

\section{Results}\label{21}

\begin{figure*}[t]
\includegraphics{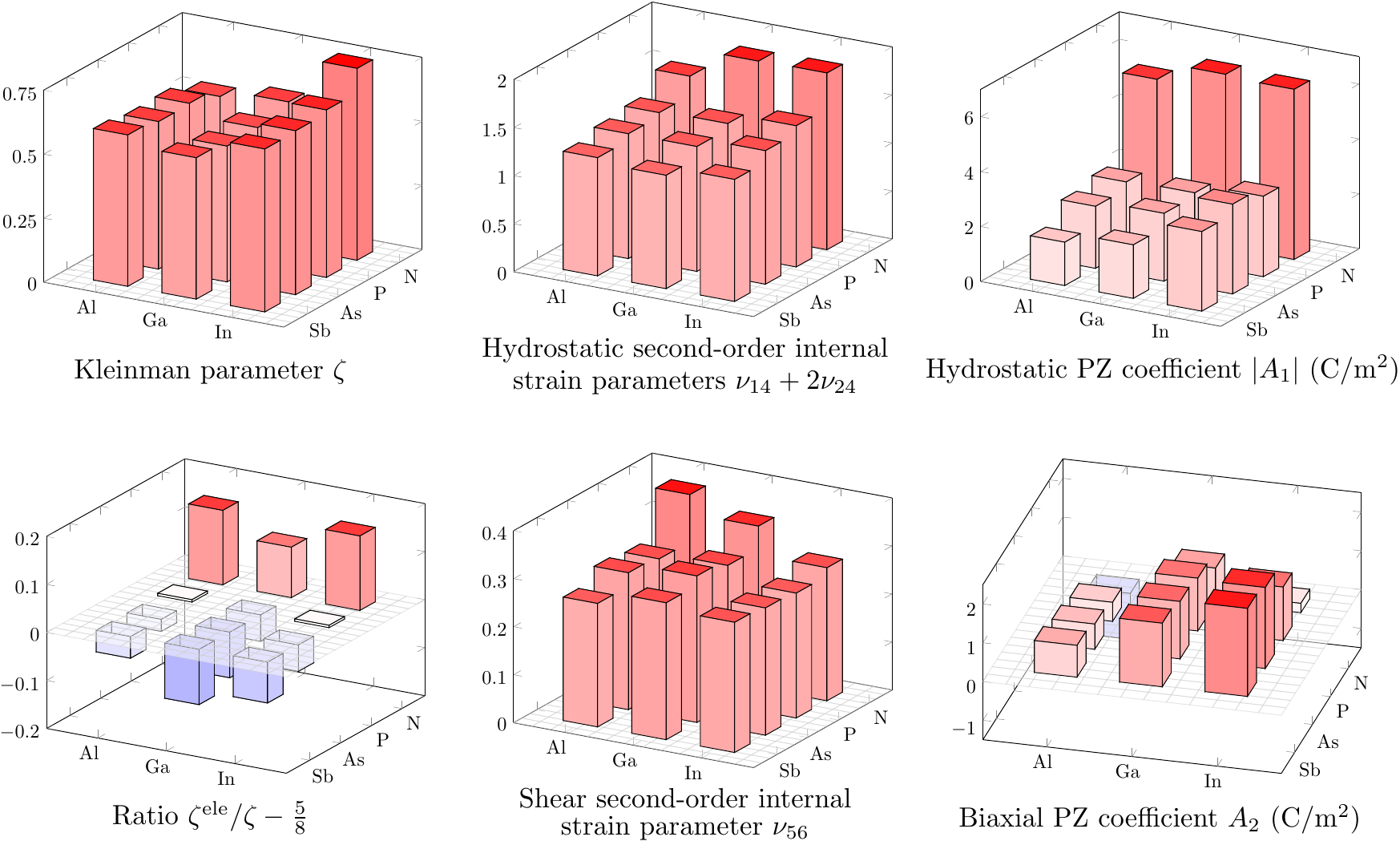} 
\caption{(Color online) Bar chart with the comparison of different
internal strain parameters and PZ coefficients for the different
III-V compounds.}
\label{18}
\end{figure*}

All our results for first- and second-order PZ coefficients of the
12 ZB binary compounds \{Al,Ga,In\}-\{N,P,As,Sb\} are summarized in
Table~\ref{03}, where we present separately the ionic and electronic
contributions to $e_{14}$, $B_{114}$, $B_{124}$, $B_{156}$, $A_1$ and $A_2$.
A number of related results are presented in Fig.~\ref{18}, and will be
discussed in more detail here and in the next section.
While trends in $B_{114}$ and $B_{124}$ are difficult
to extract,~\cite{beya_2011} the magnitude of
$A_1$ shows a clear decrease with increasing unit cell
volume and group~V row, as shown graphically in Fig.~\ref{18}.
By contrast, $A_2$ increases for fixed cation with increasing group~V row.
As mentioned above, Table~\ref{03} presents the ionic
and electronic contributions to the different coefficients separately.
It is readily clear from the numbers presented, in particular for 
$e_{14}$, that the PZ response in ZB arises
as the result of two much larger delicately balanced contributions:
the ionic and electronic parts.
These results are further discussed in Sec.~\ref{16}.

Second-order PZ coefficients for the ZB III-V family, except
the nitrides, have already been obtained in the DFT frame by Beya-Wakata
\textit{et
al.} using the LDA.~\cite{beya_2011} Our
results in Table~\ref{03} are overall very similar for the linear PZ
coefficients of the different compounds with wider band gap, but less so
for the second-order coefficients, especially for the narrow gap materials.
This disagreement is not surprising since the LDA predicts a metallic state
for some of these compounds (e.g. GaSb, InAs and InSb) which is not
compatible with the existence of a macroscopic electric polarization
in the material. Two initial conclusions can be drawn
from this comparison and from earlier work on WZ nitrides:~\cite{caro_2013b}
i) LDA and/or other DFT approximations can be sufficient to study linear
PZ polarization when a positive band gap is correctly predicted
and ii) when both schemes predict a positive gap,
the main differences between standard DFT implementations and a hybrid
functional approach arise from the disagreement in the predicted internal
strain. This second consideration also accounts for the disagreement
between LDA-DFT and HSE-DFT in the case of the spontaneous polarization of WZ
nitrides, which arises directly from the disagreement regarding the WZ internal
parameter $u$.~\cite{caro_2013b} The HSE approach has
been shown to improve upon Kohn-Sham DFT for a wide range of semiconductors
regarding the description of not only band gaps, but also lattice parameters
and elastic properties.~\cite{heyd_2003,heyd_2004,paier_2006,caro_2012c}
Therefore we expect also the description of internal strain to be superior
when using the HSE functional.

\section{Discussion and interpretation of results}\label{22}

In the following we discuss the results obtained in this work and
interpret them using a simple physical approach.
Section~\ref{15} briefly comments on separating the system into ionic
and electronic parts and the significance of defining a
reference starting point.
Sections~\ref{16} and \ref{17}
then consider the results obtained for the first- and second-order
piezoelectricity, respectively, and their interpretation
within the context of the present model.

\subsection{General considerations}\label{15}

One of the most striking features of piezoelectricity in the ZB III-Vs
is that, except for the highly ionic nitrides, the second-order PZ
response is very large if compared to the first-order contribution. This
assertion becomes extreme in the case of AlP and InP, for which the sign
of the effective PZ coefficient as a function of hydrostatic
strain $\tilde{e}_{14} = e_{14} + A_1 \text{Tr}(\epsilon)$ would
\textit{reverse} its sign for as little as $\approx 0.2\%$ tensile
hydrostatic strain. As a matter of fact, Beya-Wakata \textit{et
al}.~\cite{beya_2011} found a negative linear PZ coefficient for AlP,
while our value in Table~\ref{03} is positive. To picture the meaning
of this change in sign of the effective PZ coefficient one can imagine
a ball-and-stick model in which cations and anions are represented by
point charges of opposite sign. A positive $\tilde{e}_{14}$ corresponds
to a positively charged cation and a negatively charged anion, which
is found to be the case for the nitrides, implying in the ball-and-stick
model a valence electronic charge (to be added to the bare positive core
charge) of $< 3$ electrons on the cation and $> 5$ electrons on
the anion.

This first-order result is then consistent with Harrison's model. In
this model the bond polarity depends on the relative magnitude of the
covalent interaction $U$ and the difference $C$ in the atomic orbital
self-energies on the neighboring atomic sites. The total number
of electrons on the cation site increases as the magnitude of the
ratio of $U$/$C$ increases. The large electronegativity of N atoms
compared to the other group~V elements then leads to a smaller ratio
of $U$/$C$ for the nitrides, consistent with there being $> 5$ electrons
on the N sites in the ball-and-stick model.~\cite{harrison_1989}
We note however that the calculated negative values of $A_1$ are contrary
to what would be predicted using the Harrison model, where the magnitude
of $U$ decreases with increasing bond length, leading to a predicted
flow of charge from the cation to the anion with increasing volume.
However, the negative value of $A_1$ describes the opposite behavior --
transfer of charge from the anion to the cation with increasing volume.

\begin{figure}[t]
\centering
\includegraphics{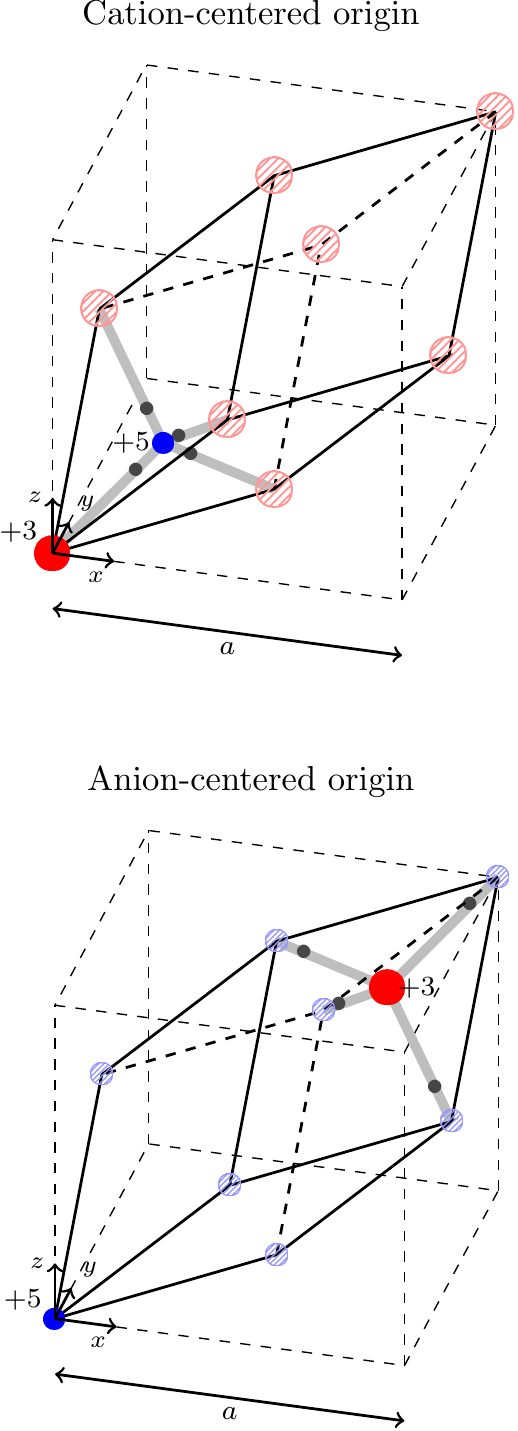}
\caption{(Color online) Zinc-blende lattice primitive
unit cells showing the underlying cubic symmetry in both
cation-at-origin (top)
and anion-at-origin (bottom) conventions. Large (red) balls
represent cations, with core charge $+3$,
while small (blue) balls represent anions, with core charge $+5$.
Solid balls are cores contained in the unit cell of the figure and
shaded balls are cores contained in neighboring unit cells. The
dark dots placed along the bonds represent (spin-degenerate) Wannier
centers of charge, with charge $-2$.}
\label{05}
\end{figure}

This disagreement in the direction of charge transfer may reflect factors
omitted from the Harrison model, including electron correlation effects.
Also, in the context of the modern theory of polarization,
a representation in terms of point dipoles lacks rigorous support.
Any attempt
to achieve a discrete representation of the system as an ensemble of
point charges requires to decouple ionic and electronic contributions,
with the point electrons given by the Wannier centers of
charge.~\cite{vanderbilt_1993} For
spin-polarized calculations the latter are assigned charge $-1$, while
in the case of spin degeneracy they are assigned charge $-2$. For the
III-Vs, the cation has core charge $+3$ and the anion has core charge $+5$.
When semicore $d$ states are considered for Ga and In, their core charges
become $+13$; however in our discussion
we will assume for simplicity that the corresponding
extra valence electrons are localized around the core giving an
effective core charge of $+3$. Then for ZB III-Vs, in the spin-degenerate
case, there are four centers of charge with charge $-2$ each, corresponding
to
the overall 8 valence electrons. This is depicted in Fig.~\ref{05}, for both
the case where the cation is chosen to be located at the origin of the
unit cell, in which we shall work preferentially since it is the
standard representation for ZB, and the case where the anion
is located at the origin.
Both representations must be (and are) equivalent for computing the
polarization. For convenience, the four Wannier centers of charge can
be further combined into their center of mass, which is then assigned
charge $-8$.

\subsection{First-order internal strain and piezoelectricity}\label{16}

The \textit{ionic} dipole moment in the cation-at-origin representation
is given by the relative position of the anion with respect to the
cation, which is kept fixed:
\begin{align}
\textbf{d}^\text{ion} = 5 e \left[ ( \mathbbm{1} + \epsilon)
\textbf{r}_0 + \textbf{t} \right].
\label{06}
\end{align}
Here, $\textbf{r}_0$ is the position of the anion before strain, and
\textbf{t} is the anion's internal strain, which for ZB is given,
\textit{to first order},
in terms of the Kleinman parameter
$\zeta$:~\cite{keating_1966,caro_2012c,caro_2013}
\begin{align}
\textbf{t}^{(1)} = -\frac{\zeta a}{4} \left(
\begin{array}{c}
\epsilon_4 \\ \epsilon_5 \\ \epsilon_6
\end{array}
\right).
\label{11}
\end{align}
The appearance of a strain-dependent part in \eq{06} is a consequence
of the choice of origin to compute the dipole moment,
and was already discussed in the seminal paper
of Vanderbilt and King-Smith.~\cite{vanderbilt_1993} Alternatively,
if we had placed the origin to compute the dipole moment at a position
exactly 5/8 along the vector connecting the cation and the anion (keeping
the same convention for the unit cell),
strain dependence would have been removed:
\begin{align}
\textbf{d}^\text{ion} = & 5 e \left[ ( \mathbbm{1} + \epsilon)
\frac{3\textbf{r}_0}{8} + \frac{3\textbf{t}}{8} \right]
+ 3 e \left[ ( \mathbbm{1} + \epsilon)
\frac{- 5\textbf{r}_0}{8} \right]
\nonumber \\
= & \frac{15e}{8} \textbf{t},
\end{align}
reaching the result that the ionic dipole moment in ZB depends only on
internal strain. Therefore any dependence on strain of this value must
arise from the dependence of \textbf{t} on strain.
Since the result must be independent of the choice of origin, the ionic dipole
moment $\textbf{d}^\text{ion}$
can be obtained as in \eq{06} and then the strain-dependent artifact
subtracted. Therefore, in the cation-at-origin convention, using
Eqs.~(\ref{06}) and (\ref{11}) we can express the
linear \textit{ionic} PZ coefficient as
\begin{align}
e_{14}^\text{ion} = - \frac{5 e}{a^2} \zeta,
\end{align}
where the $a^2$ term appears after dividing by the unit cell's volume
$V = a^3 / 4$.

Conversely, within this same convention, the linear \textit{electronic}
PZ coefficient can be expressed as
\begin{align}
e_{14}^\text{ele} = \frac{8 e}{a^2} \zeta^\text{ele},
\label{14}
\end{align}
which in turn serves as the definition for the ``electronic
Kleinman parameter'' $\zeta^\text{ele}$. Note that the strain-dependent
artifact for the electronic part arises in the same fashion as in \eq{06}
if Wannier centers are used, or arises
as the so-called ``quantum of polarization''
if the computation is done directly from the Berry phase.~\cite{resta_2007}

Finally the total linear PZ coefficient is given by
\begin{align}
e_{14} = e_{14}^\text{ion} + e_{14}^\text{ele} = \frac{e}{a^2}
(8 \zeta^\text{ele} - 5 \zeta).
\label{07}
\end{align}
Within the anion-at-origin convention, the ionic and electronic
contributions would have been given by $3 e \zeta / a^2$ and
$8 e (\zeta^\text{ele} - \zeta) / a^2$, respectively, thereby,
as expected, leaving the total PZ coefficient unaltered.

Equation~(\ref{07}) provides an extremely useful and intuitive
insight to the origin of the PZ effect in ZB III-V semiconductors.
A net dipole moment in the unit cell appears as a consequence of the loss
of centrosymmetry and can be expressed
in terms of the internal strain parameters
(see also Ref.~\onlinecite{caro_2013b}
for an equivalent treatment in wurtzite).
The magnitude of this PZ effect is then given by the ability (or inability)
of the Wannier centers of charge to follow the movement of the anion.
This idea was already highlighted by Vanderbilt:~\cite{vanderbilt_2000}
``\textit{the proper
piezoelectric response is identically zero for a homogeneous \emph{[strain]}
deformation of both the ionic positions and the Wannier
centers}''. Rearranging \eq{07} as follows allows to emphasize this idea
even further:
\begin{align}
e_{14} = \frac{e \zeta}{a^2} \left( 8 \frac{\zeta^\text{ele}}{\zeta} - 5
\right),
\label{08}
\end{align}
where the ratio $\zeta^\text{ele} / \zeta$ characterizes the ability of
the Wannier centers of charge to follow the anion core.
$\zeta^\text{ele} / \zeta = 1$ corresponds to these centers following the
core exactly, that is,
it depicts an ideal ionic picture where the charge of the
anion can be effectively expressed as its oxidation number ($-3$ in this
case). $\zeta^\text{ele} / \zeta = 0$ corresponds to the Wannier centers
fully ignoring the movement of the anion core and remaining at their
original positions in the lattice (plus a strain transformation). This
situation would interestingly lead to a larger PZ constant, of negative
sign, indicating reversed ionicity, i.e. the centers of charge follow the
\textit{cation} movement exactly. For III-Vs, charge balance is achieved
when $\zeta^\text{ele} / \zeta = 5/8 = 0.625$, in which case the PZ response
of the material is zero.
The Kleinman parameter for the different compounds
is shown in Table~\ref{09} (also in graphical
form in Fig.~\ref{18}), together with the ratio
$\zeta^\text{ele} / \zeta$, where $\zeta^\text{ele}$ is obtained
from \eq{14}. All the compounds with
$\zeta^\text{ele} / \zeta > 0.625$ present regular ionicity and a positive
PZ coefficient (basically
the nitrides), while the rest present reversed ionicity, with a net
positive charge on the anions. The proximity
of the value of $\zeta^\text{ele} / \zeta$ to 0.625 for
many of these materials then
explains the importance of the quadratic terms in computing
the PZ response of the III-Vs, and even in determining its sign.

For comparison,
we have computed the numbers for an extreme case of ionicity, ZB NaF, a
I-VII compound for which $\zeta = 0.921$ and
$\zeta^\text{ele} / \zeta = 0.982$, that is the Wannier centers of charge
follow the anionic core almost exactly. A Kleinman parameter $\zeta$
close to 1 is also indicative of high ionicity because it implies the
response of the asymmetric unit to strain is to preserve the tetrahedral
bond lengths rather than the bond angles.~\cite{kleinman_1962}

We see from Table~\ref{09} and Fig.~\ref{18} that the Kleinman parameter
has a relatively small variation across the different III-V compounds
considered, although its value does tend to be slightly higher in the
In-V compounds, reflecting that the bond-bending force constants are
relatively weaker in the In-V compounds compared to the Al- and
Ga-containing compounds. We also note that
the PZ coefficient of NaF (and other I-VIIs) is not given by \eq{08}, but by
$e_{14} = \frac{e \zeta}{a^2} (8 \frac{\zeta^\text{ele}}{\zeta} - 7)$
instead. This analysis allows to establish the maximum linear PZ coefficient
that can exist for the different families of binary
ZB compounds as a function of their lattice parameter as
$e_{14}^\text{max} = \frac{X e}{a^2}$, where $X$ is the absolute value of
the anion's oxidation number. In practice, $e_{14}$ is only close to
this value for the I-VII compounds.

\begin{table}[t]
\caption{Kleinman parameter $\zeta$ and ratio of the
``electronic Kleinman parameter'' $\zeta^\text{ele}$
with respect to $\zeta$ for all the III-Vs.}
\begin{ruledtabular}
\label{09}
\begin{tabular}{| r | c c c | r | c c c |}
\hline
 & \multicolumn{3}{c |}{\parbox{3.3cm}{Kleinman parameter $\zeta$}} & \parbox{.75cm}{\hspace{0.65cm}}
 & \multicolumn{3}{c |}{\parbox{2.4cm}{Ratio $\zeta^\text{ele} / \zeta$}}
\\
 & \parbox{1.1cm}{Al} & \parbox{1.1cm}{Ga} & \parbox{1.1cm}{In} &
 & \parbox{.8cm}{Al} & \parbox{.8cm}{Ga} & \parbox{.8cm}{In}
\\ \hline
N  & 0.539 & 0.568 & 0.749 & N  & 0.776 & 0.726 & 0.779 \\
P  & 0.580 & 0.535 & 0.654 & P  & 0.631 & 0.572 & 0.631 \\
As & 0.576 & 0.530 & 0.640 & As & 0.601 & 0.527 & 0.574 \\
Sb & 0.591 & 0.552 & 0.636 & Sb & 0.577 & 0.510 & 0.539
\\ \hline
\end{tabular}
\end{ruledtabular}
\end{table}

Another important consideration that becomes
immediately clear from \eq{08} is the
volume effect present in piezoelectricity, whereby the lattice parameter
enters the expression of the PZ coefficient as $1/a^2$. This factor varies
for the ZB III-V compounds between 0.023~\AA$^{-2}$ for InSb and 0.052~\AA$^{-2}$
for AlN, so that at equal dipole moment per atom pair,
this volume effect would enhance the PZ coefficient
of AlN by more than a factor of 2 compared to InSb.

\begin{figure*}[t]
\parbox[t]{0.49\linewidth}{\hrule height 0pt width 0pt
\includegraphics{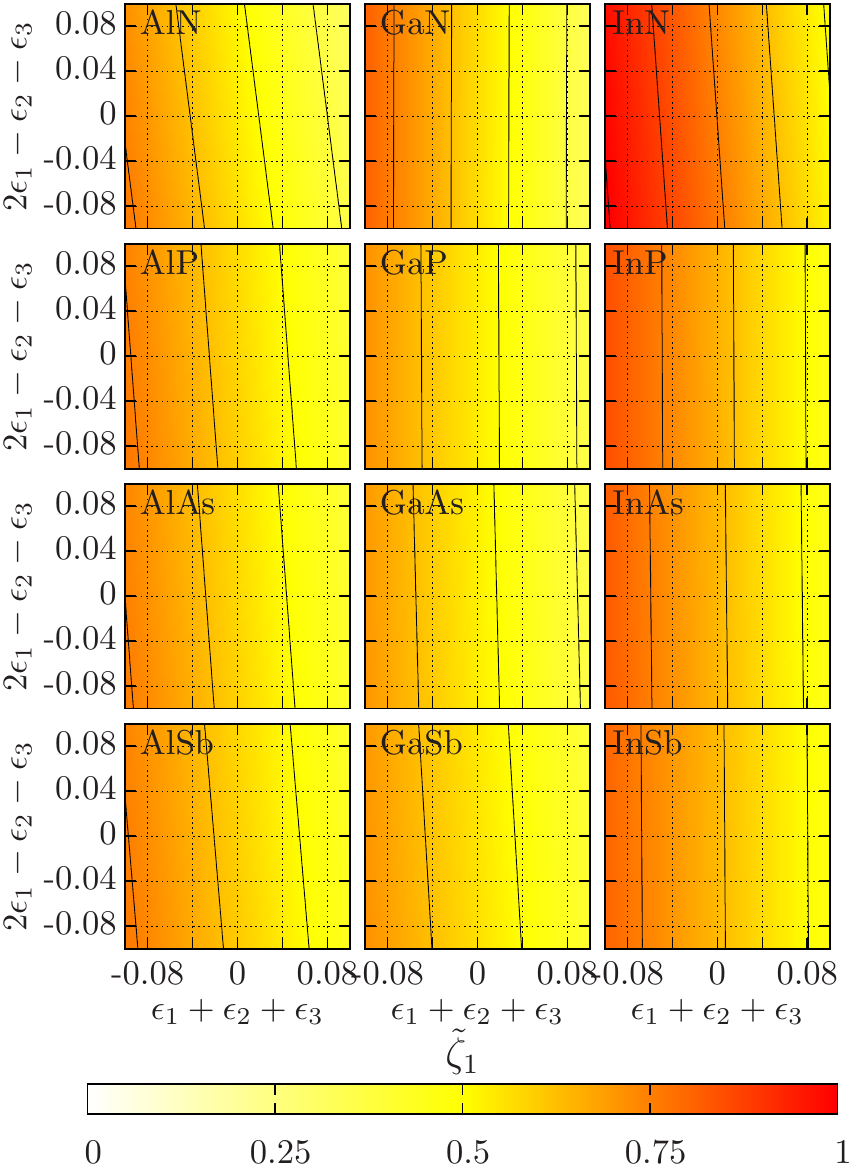}}
\parbox[t]{0.49\linewidth}{\hrule height 0pt width 0pt
\includegraphics{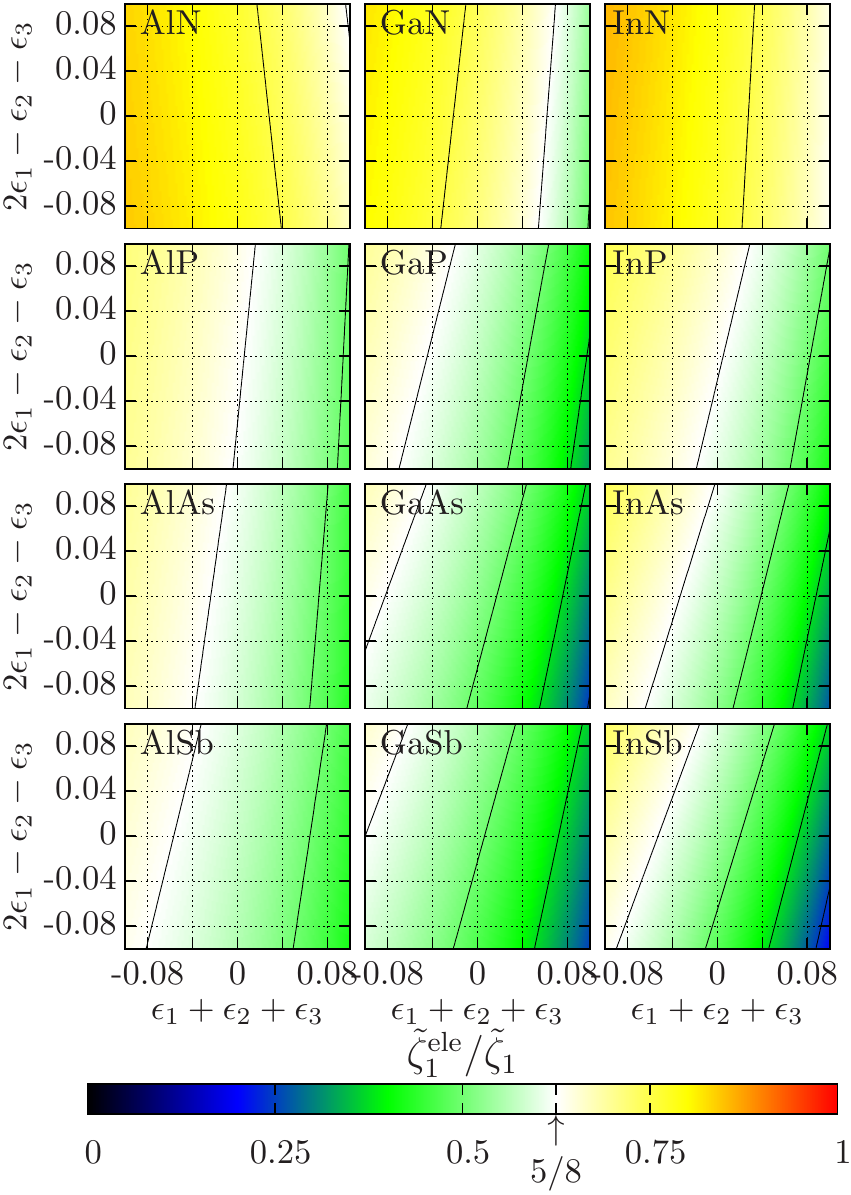}} 
\caption{(Color online) Evolution as a function of hydrostatic and biaxial
strain of the strain-dependent Kleinman parameter $\tilde{\zeta}_1$ and
$\tilde{\zeta}_1^\text{ele} / \tilde{\zeta}_1$ ratio for all the ZB
III-V binaries. The distance between adjacent contour lines is 0.125
in the color-coded scale.}
\label{12}
\end{figure*}

\subsection{Second-order internal strain and piezoelectricity}\label{17}

The second-order internal strain parameters in ZB, which we
call $\nu_{ij}$, can be obtained in
a similar way to the Kleinman parameter~\cite{keating_1966} and the wurtzite
internal strain parameters,~\cite{camacho_2010,caro_2012c} expanding the
expression of the interatomic energy to third-order in strain
and evaluating the different terms:
\begin{align}
\textbf{t}^{(2)} = &
a \nu_{14} \left(
\begin{array}{c}
\epsilon_1 \epsilon_4 \\ \epsilon_2 \epsilon_5 \\ \epsilon_3 \epsilon_6
\end{array}
\right) 
+
a \nu_{24} \left(
\begin{array}{c}
(\epsilon_2 + \epsilon_3) \epsilon_4 \\ (\epsilon_1 + \epsilon_3) \epsilon_5
\\ (\epsilon_1 + \epsilon_2) \epsilon_6
\end{array}
\right) 
\nonumber \\
& +
a \nu_{56} \left(
\begin{array}{c}
\epsilon_5 \epsilon_6 \\ \epsilon_4 \epsilon_6 \\ \epsilon_4 \epsilon_5
\end{array}
\right).
\label{10}
\end{align}
Equation~(\ref{10}) can also be expressed in terms of hydrostatic and
biaxial strain:
\begin{align}
\textbf{t}^{(2)} = &
a \nu_1 \text{Tr}(\epsilon)
\left(
\begin{array}{c}
\epsilon_4 \\
\epsilon_5 \\
\epsilon_6 \\
\end{array}
\right)
+ a \nu_2
\left(
\begin{array}{c}
\epsilon_{\text{b},1} \epsilon_4 \\
\epsilon_{\text{b},2} \epsilon_5 \\
\epsilon_{\text{b},3} \epsilon_6 \\
\end{array}
\right)
\nonumber \\
& + a \nu_{56}
\left(
\begin{array}{c}
\epsilon_5 \epsilon_6 \\
\epsilon_4 \epsilon_6 \\
\epsilon_4 \epsilon_5 \\
\end{array}
\right).
\label{25}
\end{align}
It is unsurprising that Eqs.~(\ref{10}) and (\ref{25}) resemble
the expressions of the second-order
PZ polarization [cf. Eqs.~(\ref{01}) and (\ref{24}), respectively].
We will not report their values for all the III-Vs here.
They can however be computed from the second-order \textit{ionic} PZ
coefficients
in Table~\ref{03} in a similar manner to the linear one:
\begin{align}
\nu_{14} = & \frac{a^2}{4} \frac{B_{114}^\text{ion} + e_{14}^\text{ion}}{5 e},
\qquad
\nu_{24} = \frac{a^2}{4} \frac{B_{124}^\text{ion} + e_{14}^\text{ion}}{5 e},
\nonumber \\
\nu_{56} = & \frac{a^2}{4} \frac{B_{156}^\text{ion}}{5 e},
\label{13}
\end{align}
and
\begin{align}
\nu_{1} = & \frac{a^2}{4} \frac{A_{1}^\text{ion} + e_{14}^\text{ion}}{5 e},
\qquad
\nu_{2} = \frac{a^2}{4} \frac{A_{2}^\text{ion}}{5 e},
\end{align}
where the term involving $e_{14}^\text{ion}$ in the different $\nu$s comes
from the change in volume due to the hydrostatic strain.
Looking at the coefficients for $A_{1}^\text{ion}$, 
$A_{2}^\text{ion}$ and $B_{156}^\text{ion}$ in Table~\ref{03},
and taking the scaling factor $a^2$ from \eq{13} into account, it can be seen
that the second-order internal strain shows only a relatively small variation
between the different binaries, as was also the case for
the Kleinman parameter (Table~\ref{09} and the bar charts in Fig.~\ref{18}).

The equivalent expressions for the center of mass of the Wannier centers
are:
\begin{align}
\nu_{14}^\text{ele} = & \frac{a^2}{4} \frac{B_{114}^\text{ele} + e_{14}^\text{ele}}{-8 e},
\qquad
\nu_{24}^\text{ele} = \frac{a^2}{4} \frac{B_{124}^\text{ele} + e_{14}^\text{ele}}{-8 e},
\nonumber \\
\nu_{56}^\text{ele} = & \frac{a^2}{4} \frac{B_{156}^\text{ele}}{-8 e},
\end{align}
and
\begin{align}
\nu_{1}^\text{ele} = & \frac{a^2}{4} \frac{A_1^\text{ele} + e_{14}^\text{ele}}{-8 e},
\qquad
\nu_{2}^\text{ele} = \frac{a^2}{4} \frac{A_2^\text{ele}}{-8 e}.
\end{align}
Using the expression for the second-order internal strain, \eq{10}, together
with the first order one, \eq{11}, allows to construct strain-dependent
analogues of $\zeta$ and $\zeta^\text{ele}$ as a function of hydrostatic
and biaxial strain:
\begin{align}
\tilde{\zeta}_i (\epsilon) = \zeta - 4 \nu_1
\text{Tr} (\epsilon) - 4 \nu_2 \epsilon_{\text{b},i},
\nonumber \\
\tilde{\zeta}_i^\text{ele} (\epsilon) = \zeta^\text{ele} - 4 \nu_1^\text{ele}
\text{Tr} (\epsilon) - 4 \nu_2^\text{ele} \epsilon_{\text{b},i},
\label{26}
\end{align}
with
\begin{align}
\nu_1 = \frac{\nu_{14} + 2 \nu_{24}}{3}, \qquad \nu_2 = \frac{2\nu_{14} - 2\nu_{24}}{3},
\nonumber \\
\nu_1^\text{ele} = \frac{\nu_{14}^\text{ele} + 2 \nu_{24}^\text{ele}}{3},
\qquad
\nu_2^\text{ele} = \frac{2 \nu_{14}^\text{ele} - 2 \nu_{24}^\text{ele}}{3}.
\end{align}
It can be seen from the values of $A_{1}^\text{ion}$ and $A_{2}^\text{ion}$
in Table~\ref{03}
that the magnitude of  $\tilde{\zeta}_i (\epsilon)$ tends to increase
with decreasing volume ($\text{Tr} (\epsilon) < 0$), as would be expected
from the bonds becoming more rigid with decreasing volume.
By contrast, $\tilde{\zeta}_i (\epsilon)$ has a much weaker dependence
on biaxial strain, $\epsilon_{\text{b},i}$. 

We can in addition combine the two terms in \eq{26} to rewrite \eq{08}
in the following form:
\begin{align}
\tilde{e}^i_{14} = \frac{e \tilde{\zeta}_i}{a^2} \left( 8
\frac{\tilde{\zeta}_i^\text{ele}}{\tilde{\zeta}_i} - 5 \right).
\end{align}
The second-order contribution to the PZ response then depends
both on the evolution of  $\tilde{\zeta}_i$ and of
$\tilde{\zeta}_i^\text{ele} / \tilde{\zeta}_i$: if
$\tilde{\zeta}_i^\text{ele} / \tilde{\zeta}_i$ was independent
of strain then it would be impossible for $\tilde{e}^i_{14}$ to change
sign with increasing strain.
In Fig.~\ref{12} we plot the
variation of the strain-dependent parameter $\tilde{\zeta}_i$ (left)
and the strain-dependent ratio
$\tilde{\zeta}_i^\text{ele} / \tilde{\zeta}_i$ (right) for all the
ZB III-Vs, as a function of both hydrostatic strain (change in volume)
and biaxial strain.
We see in the left hand figure for $\tilde{\zeta}_i$ that contours of
constant $\tilde{\zeta}_i$ are generally close to vertical, highlighting
that the hydrostatic strain has a considerably stronger effect 
than biaxial strain on the effective Kleinman parameter. 
The right figure elucidates how
strain induces a transition
(denoted by the white region in the color maps, for which
$\tilde{\zeta}_i^\text{ele} / \tilde{\zeta}_i = 5/8$)
from a negatively charged to a positively charged anion
in all the compounds except for
the nitrides (only GaN at very high tensile hydrostatic strain sees this
transition for the range shown).
For many of the compounds this transition occurs remarkably close
to the zero strain region, which highlights the importance of the
second-order PZ coefficients.
We also note that the contours of constant $\tilde{\zeta}_i^\text{ele} /
\tilde{\zeta}_i$ are close to vertical for the III-nitrides, becoming
more tilted toward the bottom right hand corner of the figure, reflecting
the trend previously noted that $A_1$ tends to decrease and $A_2$ tends
to increase with increasing compound's unit cell volume.
It is worth noting also that the nitrides retain their ionic character,
a fact that is also reflected in their relatively constant Born effective
charge when compared to other III-V compounds.~\cite{caro_2013b}

\section{Summary}\label{23}

In summary, we have calculated hybrid-functional first- and second-order
PZ coefficients for the whole family of zinc-blende III-V
semiconductors. The hybrid-functional approach allows a more accurate
evaluation of structural parameters and, more particularly, of band gaps.
Other DFT approximations predict a negative band gap for some of these
compounds which prevents the calculation of a meaningful
electric polarization.
We have reported the PZ coefficients split into their ionic and electronic
contributions and have shown how the large second-order corrections to the
first-order coefficients arise from the extremely delicate balance
between these two large components.

We have built a phenomenological model
of the PZ effect in the III-Vs that allows a straightforward interpretation
of first- and second-order effects in terms of changes in internal
strain and ionicity. While we find changes in internal strain
to be largely dominated by hydrostatic strain alone, changes in ionicity
are found to be also strongly sensitive to biaxial strain. These changes are more
pronounced as the unit cell size increases: they are smallest for AlN
and largest for InSb.

\begin{acknowledgments}
The authors acknowledge financial support from Science Foundation Ireland
(project numbers 10/IN.1/I2994 and 13/SIRG/2210), the European
Union's 7th Framework Programme DEEPEN (Grant Agreement No. 604416),
and computational resources from the Tyndall National Institute's
in-house computer cluster.
\end{acknowledgments}

\appendix*

\section{Calculation details}

\begin{figure}[t]
\centering
\includegraphics{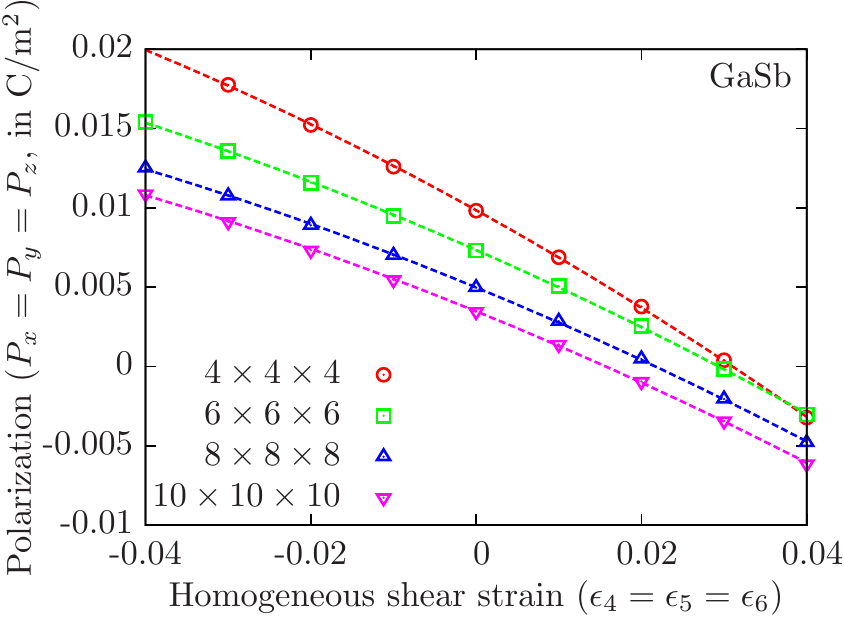}
\caption{(Color online) Convergence of the electric polarization values
in GaSb as a function of homogeneous shear strain, for different Brilloin
zone samplings.}
\label{02}
\end{figure}

The convergence of the electric polarization,
and therefore of the PZ
coefficients, is rather slow with respect to the number of \textbf{k} points
used in the calculation. This becomes problematic with the highly demanding
HSE functional, where typical computational times are one to two orders of
magnitude higher than standard implementations of Kohn-Sham DFT.
The effect of the finite $k$ sampling on the data
affects both the slope at zero strain and the curvature of the
polarization, that is first- as well as
second-order PZ coefficients.
Even for a large number
of \textbf{k} points ($10\times10\times10$) the derived values are
not fully converged.
We have found, however, a clear asymptotic behavior as the number of
\textbf{k} points grows, as can be inferred from Fig.~\ref{02} for
GaSb under homogeneous shear deformation ($\epsilon_4 = \epsilon_5
= \epsilon_6$) for different $k$ grids. The
non-zero value of the polarization at zero strain in this case can
be observed to be a numerical error due to finite $k$
sampling.~\footnote{We show the ``absolute'', in the sense of
``branch-independent'', value of the polarization. For the centrosymmetric
unstrained ZB structure this means the polarization should be ideally zero
in the absence of numerical artifacts. See
Ref.~\onlinecite{vanderbilt_2000}.} However, the error is consistent
for the different data points within the same series and therefore
the extracted PZ coefficients are independent of it.
In this context, the values of the
extrapolated PZ coefficients have been obtained
assigning different weights to the coefficients obtained with different
precision, with the lowest weight assigned to those obtained with
$4\times4\times4$ $k$ meshes and the highest weights corresponding to the
denser $10\times10\times10$ $k$ meshes.
Intermediate $6\times6\times6$
and $8\times8\times8$ meshes have also been considered.
In the LDA, an additional number of \textbf{k} points is specifically
required for materials for which a (wrong) negative band gap is predicted,
as is the case of GaSb, InN, InAs and InSb. Beya-Wakata \textit{et al}.,
for instance, used a very dense $12\times12\times12$ $k$ mesh to obtain
their LDA results.~\cite{beya_2011}
In our HSE calculations all the III-Vs studied retain a nearly correct
(as compared to experiment) positive gap, as shown in Table~\ref{04}.
This renders the matter of $k$-point density less critical.

To compensate for finite $k$-mesh sampling, we assume that the absolute
value of the numerical error $s$ in either $e_{ij}$ or $B_{ijk}$
due to finite Brilloin zone integration decays exponentially
with the number of \textbf{k} points $N_k$ used in the calculation:
\begin{align}
|s| = \alpha \text{e}^{- \beta N_k}, \qquad \alpha, \beta > 0 \,.
\end{align}
For the linear coefficient $e_{14}$, we assume the error is 50\% for the
$4 \times 4 \times 4$ mesh and 1\% for the $10 \times 10 \times 10$ mesh.
For the non-linear coefficients $B_{ijk}$ these errors are assumed to
be 100\% and 5\%, respectively. These numbers yield $\alpha = 0.653$
and $\beta = 4.18 \times 10^{-3}$ for the linear coefficients and
$\alpha = 1.227$ and $\beta = 3.2 \times 10^{-3}$ for the non-linear
ones. Then the coefficients are obtained, for each strain
branch, as a weighed average of the four available values (corresponding
to $k$-meshes $4 \times 4 \times 4$, $6 \times 6 \times 6$,
$8 \times 8 \times 8$ and $10 \times 10 \times 10$), with
weights $w (N_k) = 1 / s^2$. The final coefficient presented in
this paper is obtained as
the average of these values between the different strain branches used.

\end{document}